# Nanoscale intracellular mass-density alteration as a signature of the effect of alcohol on early carcinogenesis: A transmission electron microscopy (TEM) study

**Hemendra M. Ghimire[1], Pradeep Shukla[2], Peeyush Sahay[1], Huda Almabadi[1], Vibha Tripathi[1], Omar Skalli[3], R. K. Rao[2], Prabhakar Pradhan[1]***

[1]*Department of Physics, BioNanoPhotonics Laboratory, University of Memphis, Memphis, TN38152, USA*
[2]*Department of Physiology, University of Tennessee Health Science Center, Memphis, TN 38163, USA*
[3]*Department of Biology and Integrated Microscopy Center, University of Memphis, Memphis, TN 38152, USA*
\*Corresponding author: ppradhan@memphis.edu

**Abstract**
Alcohol consumption interferes with the functioning of multiple organ systems, causing changes in the chemistry, physiology and pathology of tissues and cellular organelles. Although epigenetic modifications underlie the development of cancer, exposure to carcinogenic chemicals, such as alcohol, can also contribute to disease development. However, the effects of chronic alcoholism on normal or pre-carcinogenic cells/tissues in different organelles are not well understood. Therefore, we herein study the effect of alcohol consumption on colonic nucleus using control and azoxymethane (AOM) and dextran sulfate sodium (DSS) treated carcinogenic mice. Previous studies showed that progression of carcinogenesis is associated with increase in the degree of intranuclear nanoscale structural disorder. In the present work, we quantify the degree of nanostructural disorder as a measure of carcinogenesis. To accomplish this, transmission electron microscopy (TEM) imaging of respective colonic epithelial cell nuclei are used to construct disordered optical lattices, and the properties of nanoscale disorder are then studied by analyzing the inverse participation ratio (IPR) of the spatially localized eigenfunctions of these optical lattices. Nanoscale structural disorder strength, as a marker of cancer progression, is measured in the length scale of 10 – 75 nm. Results show no significant visible effect in nanoscale structural changes on colon cell nuclei from alcohol exposure. However, alcohol was found to act as an enhancer of nanoscale disorder in precancerous cells and, hence, carcinogenic processes. To the best of our knowledge, this is the first study to quantify the effect of alcohol on early carcinogenic biological cells, using mesoscopic condensed matter physics.

## 1. Introduction

Alcohol consumption is considered the third leading risk factor for premature death and disability in the world [1], and the harmful use of alcohol is the component cause of more than 200 diseases [2]. Nevertheless, alcohol consumption is distinctly prevalent worldwide [3]. Similar to alcoholism, the global epidemic of cancer continues to be one of the leading causes of death around the world [4,5]. It has recently been shown that alcohol has finite effects on cancerous growth. Several oncological studies have suggested that alcohol enhances the aggravation rate of malignancies in some organs [6]. However, a detailed understanding of its organ-specific effects in early carcinogenesis is still lacking.

Recent optical spectroscopic microscopy studies have suggested that the progression of cancer in the early stage is accompanied by nanoscale morphological or nanoarchitectural alterations inside cell nuclei that precede histological abnormalities, beginning with minute rearrangements at the molecular level, such as DNA methylation, followed by genomic and proteomic alteration [7,8]. Such alteration in nuclear nanoarchitecture or morphology of nucleus results in nanoscale spatial mass-density fluctuations associated with the length scales of the principal cellular building blocks: DNA, RNA, proteins, and lipids [9]. The nucleus, the probing site in our study, is a critical organelle where task-oriented organized chromatin structure and other macromolecular complexes for each eukaryotic cell are housed, and the process of cellular response against toxicity, such as alcohol, has major effect [10-11].

Conventional visible light microscopy has been widely used to characterize biological cells/tissue in carcinogenesis [11]. However, the ability of such microscopy techniques to detect changes at the nanoscale is impeded by their diffraction-limited resolution (max ~200 nm via confocal). Yet, a comprehensive understanding of morphological changes in early carcinogenesis is critical, and, as such, appropriate quantification of structural changes well below ~100nm, the dimension of building blocks, is required. Transmission electron microscopy (TEM), on the other hand, has nanoscale resolution and is a vital tool for imaging nano- and submicron structures [12,13,14]. So far, only visual inspection of TEM nuclear images has been used for qualitative detection and inspection of nanoscale signatures. However, quantitative analysis of cellular morphology is required for an accurate estimation of cellular changes, and this aspect has not been well studied or understood, with some notable exceptions [15]. Cells are heterogeneous media associated with multiple spatial correlation length scales, making it difficult to efficiently quantify disorder strengths. To effectively calculate disorder, an indirect method is applied whereby nanoscale disorder properties of a sample are analyzed by the effective light localization properties of cellular media. To do this, we first constructed a mass-density matrix array based on TEM imaging, followed by construction of a refractive index matrix. TEM transmission intensity is proportional to the charge density in the cell, and refractive index depends on the electric polarization properties of the cell. Again, charge density and electric polarization are linearly dependent on the mass-density of a cell. Therefore, TEM intensity and the refractive index are linearly proportional. Once the refractive index matrix is created from the TEM image, we solve Maxwell's wave equation with closed boundary condition to obtain eigenvalues of the system. The inverse participation ratio (IPR) technique is then used to calculate the participation of each eigenfunction of the electric field vector. Finally, the degree of structural disorder of the system can be obtained by averaging all IPR values of the sample.

In this study, we evaluated the effects of alcohol on normal colon cells and those from a mouse model of colon cancer to examine the nanoscale structural properties of cells using transmission electron microscopy (TEM). In particular, we studied TEM images of colonic epithelial cell nuclei taken from control mice and an azoxymethane (AOM)/dextran sulfate sodium (DSS) mouse model of colon cancer without and with chronic ethanol feeding, respectively. From these, four different types of colonic epithelial cell nuclei were considered: i) Control (normal), ii) Control treated with ethanol (Control+EA), iii) AOM/DSS (AOM combined with DSS treatment), and finally, iv) AOM/DSS mice treated with ethanol (AOM/DSS+EA). Nanoscale structural disorder was calculated using the IPR technique, as noted above and as described in detail below.

## 2. Theoretical Framework:
(i) **Construction of optical refractive index array of a cell based on TEM imaging:** Biological cells are considered to be weak dielectric (or refractive index) media with refractive index varying from 1.38-1.5 [30]. Furthermore, several studies have shown strong correlation between nuclear optical refractive index (n) and local mass-density ($\rho$) of intracellular macromolecules, such as double-stranded DNAs, RNAs, aggregated chromatin, and bound proteins [31]. The refractive index of biological media can be expressed as $n = n_o + \alpha\rho$, where $n_o$ is the refractive index of the medium surrounding a scattering structure, $\rho$ is the local concentration of solids, and $\alpha$ is a proportionality constant with value nearly equal to 0.18 for most scattering substances found in living cells [16,17,18]. Therefore, if TEM imaging is performed through an ultra-thin biological sample, then it can be considered that the transmission of the contrast agent by the cell is linearly proportional to the total mass present in the thin cell voxel [32]. TEM intensity at any voxel point (x,y) is $I_{TEM}(x,y)$ and can be expressed as

$$I_{TEM}(x, y) \propto M(x, y), \qquad (1)$$

where M(x,y) is the mass-density inside the voxel at spatial position (x,y). The mass-density inside a voxel is, in turn, proportional to the refractive index n(x,y) inside the voxel, as

$$n(x, y) \propto M(x, y). \qquad (2)$$

From the above two relationships in (1) and (2), we can write the relationship between TEM local intensity and refractive index of corresponding point as follows:

$$n(x, y) \propto I_{TEM}(x, y), \qquad (3)$$

considering $n(x, y) = n_0 + dn(x, y)$, where $n_o$ is the mean refractive index, and dn(x, y) is the fluctuating part. Based on $I(x, y)_{TEM} = I_{0TEM} + dI_{TEM}$, where $I_{0TEM}$ is the mean intensity of the TEM images and $dI_{TEM}$ (x,y) is the fluctuating part, we can construct an optical lattice with an effective optical potential based on the TEM images. Accordingly, the potential in the $i^{th}$ optical lattice, $\varepsilon_i$, can be defined as

$$\varepsilon_i \propto \frac{\Delta n(x, y)}{n_0} = \frac{\Delta I_{TEM}}{I_0}. \qquad (4)$$

(ii) **Tight binding model (TBM) Hamiltonian and eigenvalue calculations:** Using the effective optical potential of the lattice system described above, we determine the Hamiltonian of the optical lattice system by Anderson's tight binding model (TBM). The TBM model is considered a good model for describing single optical states for systems of any geometry and disorder [19,20]. Only one optical state per photon per lattice site is considered, and interlattice site hoppings are restricted to the nearest neighbors, i.e., tight binding. Such Hamiltonian can be written as

$$H = \sum_i \varepsilon_i |i\rangle\langle i| + t \sum_{\langle ij \rangle} (|i\rangle\langle j| + |j\rangle\langle i|), \qquad (5)$$

where |i> and |j> are the optical wave functions at the $i^{th}$ and $j^{th}$ disordered lattice sites, respectively, <ij> indicates the nearest neighbors, and t is the overlapping integral between sites i and j.

(iii) **Inverse Participation Ratio (IPR) technique to characterize the degree of nanoscale structural disorder:** Considering the fluctuating part of the refractive index n(x,y) relative to its average background, as the rescaled potential, and entering the value of $\varepsilon_i(x,y)$, we obtain the eigenfunctions of the above Hamiltonian. We can define the average IPR value over a sample *(<IPR>$_{sample}$)* for an optical lattice of size *LxL*, where the total number of eigenfunctions is N=$(L_a)^2$ [$L_a$ = L/a (lattice size), a=dx=dy], as follows:

$$\langle IPR(L) \rangle_{Sample} = \frac{1}{N} \sum_{i=1}^{N} \int_0^L \int_0^L E_i^4(x, y) dx dy \propto dn \times L_c, \qquad (6)$$

where $E_i$ is the $i^{th}$ eigenfunction of the Hamiltonian of optical sample size LxL.

Mathematically, IPR represents the lowest value of an extended, or delocalized, eigenfunction with a value of ~2.45 for a 2D lattice in the unit of inverse area. Here, deviation value linearly increases with the degree of localization, and localization increases with the increase in disorder, being linearly proportional to <IPR> for the weak disorder case. Using a detailed numerical study, *<IPR>$_{Sample}$* α *dn x L$_c$* for weak disorder has already been shown [11]. With the IPR technique approach, all heterogeneity of the medium is expressed in a single parameter, $L_{sd} = dn \times L_c$, where $dn$ and $L_c$ belong to a chosen single Gaussian white noise disorder parameter assimilating all the heterogeneity and multilength scales into a single parameter, $L_{sd}$, thus making the degree of nanoscale structural disorder easy to compare among different samples.

## 3. Methods:

**AOM/DSS-mouse model of colon carcinogenesis**: In this study, we examined in detail the effect of alcoholism on normal and precancerous colonic epithelial cell nuclei in mice. We chose colonic epithelial cells for our study because the colon plays an important role in the digestive system with high likelihood of exposure to colon cancer-causing substances. Also, colon cancer is the most common of all gastrointestinal cancers. Mouse models of cancer have been extensively used, as mice have anatomy, physiology, and genetics similar to those of humans, shorter breeding period and accelerated lifespans. We have, therefore, used the well-characterized AOM/DSS mouse model of colon cancer. Although AOM alone has high carcinogenic potential, it has been shown that combination of AOM with DSS (AOM/DSS mice) have around ~100% probability of developing colon cancer. To model chronic alcohol consumption, we used a well-characterized Lieber-DeCarli liquid diet with or without 4% (v/v) ethanol. For controls, ethanol was replaced with isocaloric maltodextrin.

Adult female mice (10-12 weeks old) were treated with a single dose of AOM (10 mg/kg BW; *i.p.*) on day zero. Colitis was induced 5 days after AOM treatment by administering DSS (3% w/v) in drinking water for five days. DSS colitis was repeated after a 15-day recovery period. Animals euthanized and colons collected on two weeks after the second course of DSS colitis. Ethanol was administered to diet during the recovery periods. On day 44, distal colons were collected from four groups of animals: i) Control, ii) Control treated with ethanol (Control+EA), iii) AOM/DSS treated mice (AOM/DSS), and iv) AOM/DSS mice treated with ethanol (AOM/DSS+EA) and processed for TEM sample preparation.

**TEM sample preparation**: To make mechanically robust, dehydrated, and electron-transparent TEM samples from structurally weak, hydrated and electron-translucent biopsied colon samples without altering spatial structure, we followed a standard protocol, as described below. (i) Fixation: Biopsied colon samples extracted from each kind of untreated/treated animal subset were primarily fixed in 0.1 M Na cacodylate buffer (pH 7.2 to 7.4), containing 2.5 percent glutaraldehyde and 2.5 percent paraformaldehyde, for more than two hours. These fixed samples were then cut into small cubes by using two razor blades without deforming tissue and transferred to a second glutaraldehyde solution. Then, specimens were washed (10 min x 3 times) with several changes of 0.1 M cacodylate buffer (pH 7.2 to 7.4) to preserve their internal structure. We post-fixed these samples with 2 percent Osmium tetroxide ($OsO_4$) in 0.1 M Na cacodylate buffer for 1-2 hours (4°C) and then rinsed with several changes of 0.1 M cacodylate buffer (pH 7.2- 7.4), which aids in fixing lipid molecules. (ii) Dehydration: Following the standard protocol, samples were now "en block" stained with aqueous Uranyl Acetate UA and dehydrated through ethanol series (50%, 70%, 80%, 90%, 95%, 100%, and 100%) every 15 minutes, after rinsing with deionized water. (iii) Embedding and Polymerization: Samples were then embedded in polymer resin containing epoxide to stabilize them sufficiently while ultrathin sectioning, followed by polymerization at 35°C for 1day, 45°C for 1 day, and, finally, 60°C for 1 day. (iv)Trimming and Sectioning: Thoroughly cleaned and concentrated samples prepared above were then sectioned with ultra-microtone (Model-123) to a thickness of 70 nm in order to make the samples electron-transparent and mechanically robust. (v) Staining: Ultrathin specimens fixed in a grating grid were next stained with uranyl acetate and lead citrate so that our sample would become electron-transparent, as well as mechanically robust. The specimens were now extremely thin for penetration by highly absorbable electrons and thus ready for TEM imaging.

TEM imaging: TEM imaging was performed using a Joel JEM-1200 transmission electron microscope system fitted with a Hamamatsu ORCA HR camera at 810 Exposure, 2 Gain and 1 Bin, for the imaging system. A high voltage 60keV electron beam as a source of illumination with a fixed magnification of 6000X was used to obtain each electron micrograph of the samples. For the statistical analysis of IPR values associated with mass-density fluctuation, 8-10 TEM images of each mouse colon cell nucleus from each of the following categories (3 mice per category) were analyzed: (i) Control, (ii) Control + EA, (iii) AOM/DSS, and (iv) AOM/DSS + EA.

## 4. Results and Discussion:

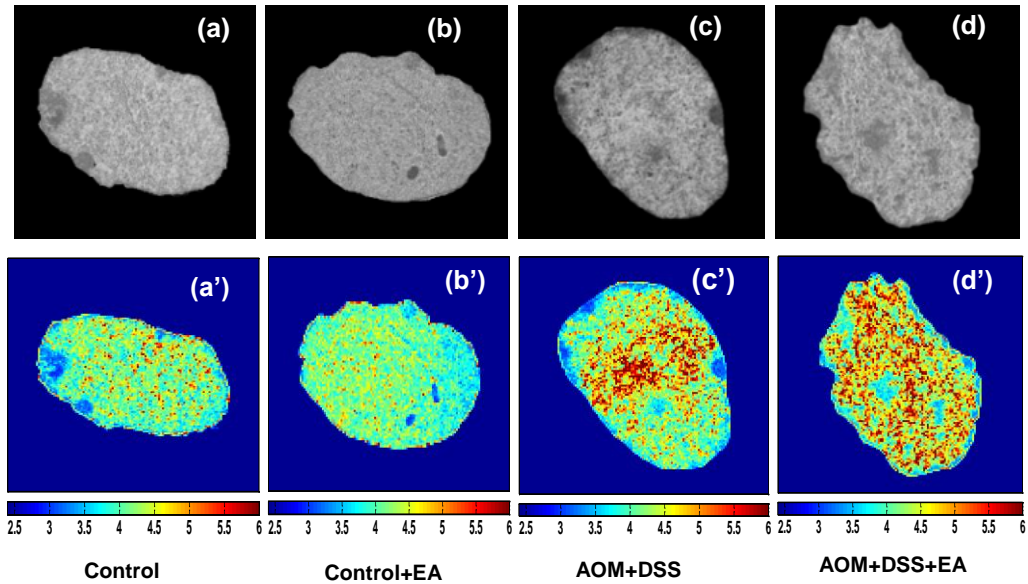

**Figure 1.** [(a), (b), (c) and (d)] are representative TEM images of nuclei taken from Control, Control + EA, AOM/DSS-treated mice and AOM/DSS + EA-treated mice, respectively. [(a'),(b'),(c') and (d')]are the corresponding $<IPR>_{Sample}$ (sizes $L \times L = 50nm \times 50nm$, resolution~2.46 nm).

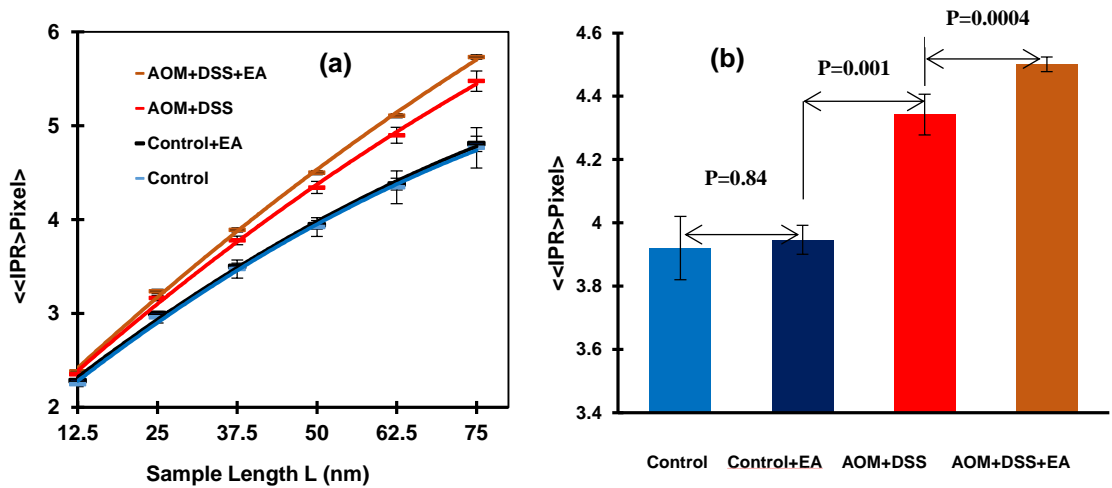

**Figure 2.** (a) Ensemble averaged $<IPR>_{Sample}$ versus sample length L (nm) plots for (i) normal colon cell nucleus, (ii) AOM plus DSS-treated cancerous colon cell nucleus, (iii) AOM plus DSS and Ethyl alcohol-treated cancerous colon cell nucleus. (b) Bar graph representation of respective mean values with length scale at pixel size *50 nm x 50* nm.

Light localization properties in the cell nuclei of all samples were evaluated and compared. Structural disorder, $L_{sd}$, in terms of IPR value, was determined for each cell at various length scales L, ranging from 12.5 nm to 75 nm. Subsequently, a mean $L_{sd}$ was obtained by averaging IPR values of all the cells from each category for the same length scales. It is interesting to point out that such length scales generally correspond to the dimension of the building blocks of the cell, e.g., DNA, RNA, or lipids. Therefore, nanostructural disorder of such ultra-scale level may correspond to early-stage deformations in these building blocks. Results obtained are presented in Figures 1 and 2. Figure 1(a), (b), (c) and (d) show representative TEM grayscale micrographs obtained respectively from colon cell nucleus of normal (Control), alcoholic (Control + EA), early colon carcinogenesis (AOM/DSS), and early colon carcinogenesis with alcohol (AOM/DSS + EA). Figure 1 (a'), (b'), (c') and (d') show their corresponding IPR images taken at the length scale of 50 nm and sample size $L \times L = 50 \times 50$ nm$^2$. It should be noted that the conventional methods of comparing intensities of grayscale images, either by visual estimation or quantitative studies, do not provide sufficient information to characterize mass-density fluctuations in the nucleus. However, the IPR images, as represented in Figures 1(a'), (b'), (c'), and (d'), clearly show significant difference in comparison to the grayscale TEM images, indicating the difference in disorder for cell nuclei treated with different cases. The intensity pattern of higher fluctuations (hot/red spots) increases from Figures 1 (a') to 1(b'), 1(c') and 1(d'), respectively. These results directly correlate with the refractive index fluctuation of the medium and, hence, the mass-density fluctuation, as described earlier. The results also suggest that the IPR images show differences, which are otherwise impossible to visualize in grayscale TEM images. Therefore, for quantitative analysis, IPR is an important parameter for nanoscale disorder and has the potential to detect early carcinogenesis in human colon cancer.

Based on the above information, IPR values of each category, representing disorder strength $L_{sd}$, were calculated for the following length scales $L$ (sample size $L \times L$): 12.5nm, 25nm, 37.5nm, 50nm, 62.5nm, and 75nm. The length scale-dependent average of $<<IPR(L)>_{sample}>$ ( $<>$_many cells) for each disorder sample were then statistically analyzed and plotted in a bar graph for one length (50nm*50nm). Figure 2(a) shows the plots of ensemble averaging of average IPR value of a sample $<<IPR(L)>_{Sample}>$ versus length $(L)$ for four different groups of cell nuclei collected from the colons of (i) Control, (ii) EA (ii) AOM/DSS, (iii) AOM/DSS+EA mice. The data show that the length scale-dependent average of $<<IPR(L)>_{Sample}>$ for each disorder sample increases with sample size and disorder fluctuation height. The decreasing slope suggests that the rate of fluctuation is going to saturate at a constant value. It also clarifies that the $<<IPR(L)>_{Sample}>$ value is highest for the AOM/DSS+EA group and the lowest for control group. For length scale of 75 nm, numerical values of $<<IPR(L)>_{sample}>$ for Control is 3.93, Control+EA is 3.94, AOM/DSS is 4.34 and AOM /DSS + EA is 4.5.

## 5. Conclusions:

We report the effect of alcohol on normal and precancerous cell nuclei by considering the strength of nanoscale structural disorder as a biomarker. The analyses of nanoscale architectural and morphological characteristics of TEM images (resolution of 2.5 nm) have shown promising results by providing a tool whereby early cancerous changes can be quantified before carcinogenesis becomes microscopically evident. Our study technique is unique in that it is based on advanced mesoscopic physics where we first convert nanoscale intracellular spatial fluctuations in mass density into spatial refractive index fluctuations via TEM imaging, followed by analysis of these refractive index fluctuations by the IPR technique. Such analysis of localized eigenfunctions derived from an optical lattice system, as derived from TEM data, enables us to quantify nanoscale morphological alterations in intracellular mass-density in a single parameter, i.e., structural disorder, $L_{sd} \sim <IPR> \alpha\ dn \times L_c$, where dn is fluctuation in refractive index with correlation length $L_c$. Based on the obtained results, the following conclusions can be drawn:

1) **Alcohol-treated control mice**: The degree of nanoscale structural disorder $L_{sd} = dn^2 \times L_c$ did not significantly change for alcohol-treated colon cell nuclei relative to control cell nuclei. This implies that alcohol may have no significant effect on normal colon cells relative to carcinogenesis.

2) **AOM/DSS-mice without chronic alcoholism**: Colon cell nuclei show significant increase in the nanoscale structural disorder in progressive carcinogenesis relative to nuclei from normal or ethanol fed mice. In particular, the value of $L_{sd}$ increases with the progression of carcinogenesis. This result is consistent with optical spectroscopy findings.

3) **AOM/DSS-mice with alcohol treatment**: Colonic epithelial cell nuclei from AOM/DSS+EA mice show higher $L_{sd}$ values compared to that in nuclei of AOM/DSS mice. In summary, alcohol by itself produce no significant effect on carcinogenic signal, but it enhances the carcinogenic process induced by AOM and DSS. These results confirm our hypothesis that no visible effect of alcohol on normal cells is found for non-carcinogenic cells, whereas in the presence of pre-carcinogenesis, chronic alcoholism acts as an enhancer of cancer progression. These basic findings also confirm that nanoscale structural disorder is a potential biomarker for cancer diagnostics and will have potential applications for cancer diagnosis and delivery of treatment regimens.

**Acknowledgements**
This work was supported by NIH-R01 and UofM (PP) and. NIH AA12307 and DK55532 ( R.K.Rao)